\documentclass[a4paper, 10pt, conference]{ieeeconf}      

\IEEEoverridecommandlockouts

\usepackage{graphicx, epstopdf}
\usepackage{cite}
\usepackage{enumerate}
\usepackage{bbm}
\usepackage{algorithmicx}
\usepackage{algorithm}
\usepackage{latexsym,bm,amsmath,amssymb,amsfonts}
\usepackage{multirow}
\usepackage{algpseudocode}
\usepackage{pifont}
\usepackage{color}
\usepackage{alltt}
\usepackage[hidelinks]{hyperref}
\usepackage{siunitx}
\usepackage{pbox}
\usepackage{picinpar}
\usepackage{url}
\usepackage{flushend}
\usepackage[latin1]{inputenc}
\usepackage{colortbl}
\usepackage{soul}
\usepackage{diagbox}
\usepackage{subfigure}

\newcommand{\RX}{\mathbf{R}}

\begin{document}

\title{\LARGE \bf Short-Term Traffic Forecasting Using High-Resolution Traffic Data}

\author{Wenqing Li$^{1}$, Chuhan Yang$^{2}$, and Saif Eddin Jabari$^{\star,3}$
	\thanks{$^{\star}$Corresponding author}
	\thanks{*This work was supported in part by CITIES (NYUAD Institute Award CG001 and Swiss Re - Quantum Cities$^{\mathrm{TM}}$ initiative).}
	\thanks{$^{1}$Wenqing Li is with the Division of Engineering, New York University Abu Dhabi, Abu Dhabi, UAE
		{\tt\small wl54@nyu.edu}}%
	\thanks{$^{2}$Chuhan Yang is with the Department of Computer Science and Engineering, Tandon school of Engineering, New York University, Brooklyn, NY, USA {\tt\small cy1004@nyu.edu}}%
	\thanks{$^{3}$Saif Eddin Jabari is with the Division of Engineering, New York University Abu Dhabi, Abu Dhabi, UAE and the Department of Civil and Urban Engineering, Tandon School of Engineering, New York University, Brooklyn, NY, USA
		{\tt\small sej7@nyu.edu}}%
}

\maketitle

\begin{abstract}
This paper develops a data-driven toolkit for traffic forecasting using high-resolution (a.k.a. event-based) traffic data.  This is the raw data obtained from fixed sensors in urban roads.  Time series of such raw data exhibit heavy fluctuations from one time step to the next (typically on the order of 0.1-1 second).  Short-term forecasts (10-30 seconds into the future) of traffic conditions are critical for traffic operations applications (e.g., adaptive signal control). But traffic forecasting tools in the literature deal predominantly with 3-5 minute aggregated data, where the typical signal cycle is on the order of 2 minutes. This renders such forecasts useless at the operations level.  To this end, we model the traffic forecasting problem as a matrix completion problem, where the forecasting inputs are mapped to a higher dimensional space using kernels.  The formulation allows us to capture both nonlinear dependencies between forecasting inputs and outputs but also allows us to capture dependencies among the inputs.  These dependencies correspond to correlations between different locations in the network.  We further employ adaptive boosting to enhance the training accuracy and capture historical patterns in the data.  The performance of the proposed methods is verified using high-resolution data obtained from a real-world traffic network in Abu Dhabi, UAE. Our experimental results show that the proposed method outperforms other state-of-the-art algorithms. 


\end{abstract}

%
\IEEEpeerreviewmaketitle

\section{Introduction}
%
%
%
%
The ability to predict future traffic is an essential component of modern intelligent transportation systems (ITS). Traffic prediction not only assists system operators to schedule interventions, but also provides travelers with routing guidance. Previous works on traffic prediction are generally based on aggregated data over certain time intervals  (no shorter than 5 minutes). Although aggregated data based prediction methods have been demonstrated to be successful, the limitation is also apparent: they may not work for the operational decisions (i.e., signal timing) that require traffic information in very short interval.  On the other hand, due to the increasing development of Internet-of-Things (IoT) technologies, especially data processing techniques and sensor technologies, high-resolution data can be stored by many traffic systems, e.g., SMART-Signal system \cite{liu2008smart} and Split Cycle and Offset Optimization Technique (SCOOT) \cite{blanc2008scoot}. Thus, it is possible to leverage the information obtained from high-resolution data to predict traffic more effectively. 

Nevertheless, accurate traffic prediction based on high-resolution data is challenging, because high-resolution data generally exhibit strong fluctuations. For example, the SCOOT system records detector occupancies every second as binary variables, where 0 represents an unoccupied sensor and 1 represents an occupied sensor.  Over a short interval, the occurrence of `0's and `1's is highly irregular.  As a result, the numbers of `0's and `1's in the high-resolution data are typically unbalanced over time. This lack of `balance' in high-resolution data presents a challenge to forecasting problems (particularly when applying traditional variants of time-series techniques): the data class that is of lower frequency tends to be overlooked by conventional prediction models, especially at times when the imbalance is pronounced. In essence, the lower frequency label is treated as an outlier.

This paper presents a matrix completion approach for traffic forecasting using high-resolution traffic data. We essentially apply techniques that have shown success in other data mining applications (e.g., image processing), where we overcome the noisy nature in the data by observing data over a longer period of time while applying techniques that can handle large problems.  We apply nonlinear kernels so that nonlinear trends in the data can be capture implicitly, which is allowed by the fast estimation techniques that we employ.  We also employ adaptive boosting (AdaBoost) where we include data from previous days to `boost' the prediction. \emph{This is a shortened version of our full paper, which includes all proofs and more extensive experimental results; we refer to \cite{li2020nonlinear} for more information.}

\section{Related work}
Data-driven prediction methodologies fall in one of two major categories: parametric approaches and non-parametric approaches. Parametric methods apply theoretical assumptions and the model parameters are calibrated using empirical data. Among parametric methods, traditional regression and filtering techniques \cite{zheng2018traffic,jabari2018stochastic,8835040}, the autoregressive (AR) family of models, including autoregressive integrated moving average (ARIMA) \cite{kumar2015short, williams2003modeling, chen2011short}, and vector autoregressive (VAR) \cite{stock2001vector,kamarianakis2003forecasting,kamarianakis2005space,kamarianakis2012real,ghosh2009multivariate} models have been widely used and demonstrated to be successful in capturing mean trends in the data but fail to capture rapid fluctuations. This implies that AR based models may not be suitable for high-resolution data.  
On the other hand, non-parametric methods do not assume a fixed model form and are typically data-driven. The basic idea behind non-parametric techniques is that they learn a general form from data and use it to predict future data. Non-parametric methods can be divided into two distinct types: non-parametric regression such as support vector regression (SVR) \cite{wu2004travel, hu2014traffic, dilip2017sparse, jabari2020sparse} and artificial neural networks (ANNs) \cite{lv2014traffic,ma2015long,mackenzie2018evaluation,ma2017learning,liu2017short}. SVR has been successfully applied to predict traffic data, e.g., flow \cite{hu2014traffic} and travel times \cite{wu2004travel}, as SVR models have powerful representation learning ability by using kernels. Alternatively, ANNs are among the first non-parametric methods that have been applied to traffic prediction, and thus there is vast literature on the subject, which extend from the simple multilayer perceptrons (MLP) \cite{lv2014traffic} to more complicated structures as generative adversarial  networks (GAN) \cite{lin2018pattern}, recurrent neural networks (RNNs) \cite{mackenzie2018evaluation}, convolutional neural networks (CNNs) \cite{ma2017learning} and the combination of RNNs and CNNs \cite{liu2017short,shi2020spatial}.

It has become well established that the higher the resolution of the data, the poorer the performance of the prediction method \cite{vlahogianni2011temporal,oh2005exploring,manual2000highway,tan2016short}. For example, the Highway Capacity Manual \cite{manual2000highway} recommends data aggregation at the 15-minute level and Tan et al. \cite{tan2016short} suggest 3 minutes as a lower bound aggregation threshold for prediction. Consequently, to our knowledge, no studies have attempted to perform short-term forecasts with high-resolution traffic data.
 
\section{Ensembled Kernelized Matrix Completion for High-resolution Data based Real-time Prediction}

In this section, the proposed \emph{ensembled kernelized matrix completion} (EKMC) algorithm is presented. We start by recasting traffic prediction problem in the matrix completion framework, and then propose a \emph{kernelized matrix completion} (KMC) algorithm which seeks to extract representative features so to achieve better prediction accuracy. Finally, we present the EKMC algorithm for traffic prediction based on the high-resolution data with unbalanced sizes. 

\subsection{Traffic Prediction as a Kernelized Matrix Completion Problem}

\subsubsection{Reformulation of Traffic Prediction in the Framework of Matrix Completion}

For traffic prediction, it is common to assume that the future data can be predicted based on the recent data.   As a result, the task of traffic prediction is to learn a mapping between the output space of the predicted data (future data) and input space of predictors (recent data). 

Suppose that $\mathbf{x}(t)\in\mathbb{R}^n$ denote a sample of a $n$-dimensional multivariate time series at the $t$-th time stamp. The recent data and predicted data w.r.t. time stamp $t$ is expressed as $\mathbf{x}_t=[\mathbf{x}(t-L+1);\cdots;\mathbf{x}(t)]\in\mathbb{R}^{nL}$ and $\mathbf{y}_t=\mathbf{x}(t+H)\in\mathbb{R}^n$ respectively, where $L$ and $H$ denotes the time lag (the length of recent data used for prediction) and prediction horizon (steps ahead to be predicted). Assume a linear relationship between the input $\mathbf{x}_t$ and output $\mathbf{y}_t$, we have
\begin{equation}
\mathbf{y}_t=\left \langle \mathbf{W},\mathbf{x}_t  \right \rangle
\label{eq:pre},
\end{equation} 
where $\mathbf{W}$ is the regression coefficient matrix that can be calculated based on training data set $\{\mathbf{x}^{tr}_t, \mathbf{y}^{tr}_t\}$. For each pair of  testing data $\{\mathbf{x}^{te}_t, \mathbf{y}^{te}_t\}$, the future sample is typically predicted  as:
\begin{equation}
\mathbf{y}^{te}_t=\left \langle \mathbf{W},\mathbf{x}^{te}_t \right \rangle
\label{eq:2}.
\end{equation}

Let us define the training matrix pair $\{\mathbf{X}^{tr}=[\mathbf{x}_t,\cdots,\mathbf{x}_{t+t_1-1}]\in\mathbb{R}^{nL \times t_1}, \mathbf{Y}^{tr}=[\mathbf{y}_t,\cdots,\mathbf{y}_{t+t_1-1}]\in\mathbb{R}^{n\times t_1}\}$ and testing matrix pair $\{\mathbf{X}^{te}=[\mathbf{x}_{t+t_1},\cdots,\mathbf{x}_{t+t_2+t_1-1}]\in\mathbb{R}^{nL \times t_2},\mathbf{Y}^{te}=[\mathbf{y}_{t+t_1},\cdots,\mathbf{y}_{t+t_1+t_2-1}]\in\mathbb{R}^{n\times t_2}\}$. The \emph{joint matrix} concatenating both training and testing data is formulated as:
\begin{equation}
\mathbf{Z}\equiv  \begin{bmatrix}
 \mathbf{Y}^{tr}&\mathbf{Y}^{te} \\ 
\mathbf{X}^{tr}&\mathbf{X}^{te} 
\end{bmatrix}\in\mathbb{R}^{(n+nL)\times (t_1+t_2)}
\label{eq:3}.
\end{equation}

In this way, the prediction of $\mathbf{Y}^{te}$ can be formulated as a matrix completion problem with $\mathbf{Y}^{te}$ unknown. As there are relationships between the input and output, termed $\begin{bmatrix}
 \mathbf{Y}^{tr}&\mathbf{Y}^{te} 
\end{bmatrix}=\mathbf{W}\begin{bmatrix}
\mathbf{X}^{tr}&\mathbf{X}^{te}
\end{bmatrix}$, $\mathbf{Z}$ is supposed to be low rank. Thus, the matrix completion of $\mathbf{Z}$ can be achieved by rank-minimization approaches \cite{jain2013low,candes2009exact}.

\subsubsection{Kernelized Matrix Completion for Traffic Prediction}
As the assumption of linear dependency between recent data and future data is idealized and uncommon in practice, we further consider a nonlinear dependency between them. 
A popular way to develop such a nonlinear mapping is to first map the input data to a higher dimensional feature space and then make a linear regression in that feature space \cite{suykens1999least}. The nonlinear relationship between the input and output is expressed as:
\begin{equation}
\begin{split}
&\mathbf{y}_t=\left \langle \mathbf{W},\phi(\mathbf{x}_t)  \right \rangle
\label{eq:1},
\end{split}
\end{equation}
where $\phi$ denotes the nonlinear function that maps the input space to the high-dimensional feature space, $\mathbf{W}$ is the regression matrix. It is worth noting that the inner product of $\mathbf{W}$ and $\phi(\mathbf{x})$ can be learned using \emph{kernel trick} without explicitly computing the map $\phi$, i.e., see \cite{smola2004tutorial,li2017linearity}. Likewise, the \emph{joint matrix} in this kernel setting is defined as:
\begin{equation}
\mathbf{Z}\equiv  \begin{bmatrix}
 \mathbf{Y}^{tr}&\mathbf{Y}^{te} \\ 
\phi(\mathbf{X}^{tr})&\phi(\mathbf{X}^{te}) 
\end{bmatrix}\in\mathbb{R}^{(n+h)\times (t_1+t_2)}
\label{eq:3},
\end{equation}
where $h \le \infty$ denotes the dimension of the feature space.

Similarly, $\mathbf{Z}$ is deemed to be low rank or be approximately low rank. Thus, the prediction problem can be formulated as a matrix completion problem and we seek to find a low rank approximation $\tilde{\mathbf{Z}}$ of the matrix $\mathbf{Z}$.

Let $P_{\Omega}$ be a binary mask that satisfies:
\begin{equation}
P_{\Omega}(\mathbf{M})=\left\{\begin{matrix}
 \mathbf{M}_{ij}&(i,j)\in \Omega \\ 
 0 & \mathrm{otherwise} 
\end{matrix}\right.
\label{eq:4}.
\end{equation}
where $\mathbf{M}$ is an arbitrary matrix. The matrix completion problem is subsequently  formulated as:
\begin{equation}
\tilde{\mathbf{Z}} = \arg \min_{\mathbf{M}} ~ \mathrm{rank}(\mathbf{M}) \quad
\mathrm{s.t.} \quad P_{\Omega}(\mathbf{M}-\mathbf{Z})=\mathbf{0}
\label{eq:5},
\end{equation}
where $P_{\Omega}$ is used to ensure that all the data are known except the testing predicted data ($\mathbf{Y}^{te}$).  
As minimizing the rank is NP hard, eq. \eqref{eq:5} is relaxed to
\begin{equation}
\tilde{\mathbf{Z}} = \arg \min_{\mathbf{M}} ~ \|P_{\Omega}(\mathbf{M}-\mathbf{Z})\|^2_F + \lambda \| \mathbf{M} \|_* 
\label{eq:7},
\end{equation}
where the nuclear norm $\|\cdot\|_*$ is a convex surrogate of the rank function, and $\lambda$ is a Lagrangian multiplier. Eq.\eqref{eq:7} is further equivalent to the following problem \cite{recht2010guaranteed},
\begin{multline}
(\tilde{\mathbf{U}},\tilde{\mathbf{V}}) \\ = \arg \min_{\mathbf{U},\mathbf{V}}  \|P_{\Omega}(\mathbf{U}\mathbf{V}^\top-\mathbf{Z})\|^2_F + \lambda(\|\mathbf{U}\|_{F}^2 + \|\mathbf{V}\|_{F}^2)
\label{eq:9},
\end{multline}
where $\mathbf{U}\in\mathbb{R}^{(n+h)\times r}$, $\mathbf{V}\in\mathbb{R}^{(t_1+t_2)\times r}$ and $r \ge \mathrm{rank}(\tilde{\mathbf{Z}})$.  We further divide $\mathbf{U}$ and $\mathbf{V}$ into two blocks each: a training block and a testing block: 
\begin{equation}
	\mathbf{U}=\begin{bmatrix} \mathbf{U}^{tr}\\ \mathbf{U}^{te} \end{bmatrix} \mbox{ and } \mathbf{V}=\begin{bmatrix}
	\mathbf{V}^{tr}\\ \mathbf{V}^{te} \end{bmatrix}.
\end{equation}
This generates the following optimization problem (termed \emph{ kernelized matrix completion}, KMC):
\begin{multline}
(\tilde{\mathbf{U}}^{tr},\tilde{\mathbf{U}}^{te},\tilde{\mathbf{V}}^{tr},\tilde{\mathbf{V}}^{te})
= \underset{\mathbf{U}^{tr},\mathbf{U}^{te},\mathbf{V}^{tr},\mathbf{V}^{te}}{\arg \min} \|\mathbf{Y}^{tr}-\mathbf{U}^{tr}(\mathbf{V}^{tr})^\top \|^2_F \\
+\|\Phi^{tr} -\mathbf{U}^{te}(\mathbf{V}^{tr})^\top \|^2_F 
+\|\Phi^{te}-\mathbf{U}^{te}(\mathbf{V}^{te})^\top \|^2_F \\ 
+\lambda( \|\mathbf{U}^{tr} \|^2_F + \|\mathbf{U}^{te} \|^2_F + \|\mathbf{V}^{tr} \|^2_F + \|\mathbf{V}^{te} \|^2_F)
\label{eq:10},
\end{multline}
where $\Phi^{tr}=\phi(\mathbf{X}^{tr})$ and $\Phi^{te}=\phi(\mathbf{X}^{te})$.

\subsection{Optimization for KMC}
The optimization of KMC may not be joint convex for all variables, however, it is convex w.r.t. each of them while keeping others fixed. Thus, coordinate descent algorithm \cite{xu2013block} is used over blocks to obtain a stationary point.

Denote the KMC objective function by $F(\RX) = F(\RX_1,\RX_2,\RX_3,\RX_4)$, where $\RX$ is the four-block matrix $\RX_1 \equiv \mathbf{U}^{tr}$, $\RX_2 \equiv \mathbf{U}^{te}$, $\RX_3 \equiv \mathbf{V}^{tr}$, and $\RX_4 \equiv \mathbf{V}^{te}$.  The block-coordinate descent solves for each of these four blocks separately in each iteration. Let $\RX^k$ denote the solution in iteration $k$, then
\begin{multline}
	\RX_1^k = \arg \min_{\RX_1} F(\RX_1, \RX_2^{k-1}, \RX_3^{k-1}, \RX_4^{k-1}) \\
	= \mathbf{Y}^{tr}\RX_3^{k-1}((\RX_3^{k-1})^\top \RX_3^{k-1} +\lambda\mathbf{I})^{-1},
\end{multline}
\begin{multline}
	\RX_2^k = \arg \min_{\RX_2} F(\RX_1^k, \RX_2, \RX_3^{k-1}, \RX_4^{k-1}) \\
=(\Phi^{tr}\RX_3^{k-1}+\Phi^{te}\RX_4^{k-1})((\RX_3^{k-1})^\top\RX_3^{k-1} \\+(\RX_4^{k-1})^\top \RX_4^{k-1}+\lambda\mathbf{I})^{-1},
\end{multline}
\begin{multline}
\RX_3^k = \arg \min_{\RX_3} F(\RX_1^k, \RX_2^k, \RX_3, \RX_4^{k-1}) \\
=((\mathbf{Y}^{tr})^\top \RX_1^{k} + (\Phi^{tr})^\top \RX_2^{k})((\RX_2^{k})^\top \RX_2^{k} \\+ (\RX_1^{k})^\top \RX_1^{k} +\lambda\mathbf{I})^{-1},
\end{multline}
and
\begin{multline}
\RX_4^k = \arg \min_{\RX_4} F(\RX_1^k, \RX_2^k, \RX_3^k, \RX_4) \\
= (\Phi^{te})^\top\RX_2^k ((\RX_2)^\top \RX_2+\lambda\mathbf{I})^{-1}.
\end{multline}
%

In each iteration, the above algorithm solves four least-squared (LS) problems and has a complexity of $O(t_1^2r)$ for each iteration.  Moreover, the number of iterations needed to obtain a reasonable solution tends to be small, since  solving for $\RX_1$, $\RX_2$, $\RX_3$, and $\RX_4$ iteratively, we can guarantee convergence to \emph{block-coordinate-wise minimizer} and the rate of convergence is sublinear, i.e., the difference between the solution $\RX^k$ in iteration $k$ and the fixed point solution $\RX^*$ is proportional to $k^{-1}$. We refer to \cite{li2020nonlinear} for a detailed development of both of these claims.


\subsection{Ensembled Kernelized Matrix Completion for High-resolution Data based Prediction}
Considering the unbalanced data size of our high-resolution data, this section proceeds to propose the \emph{ensembled kernelized matrix completion} (EKMC) method. The EKMC procedure is detailed below.


\subsubsection{Data Arrangement} 
Since traffic data exhibit periodical patterns, we include data from $T$ time steps before the current time from the recent $d$ days in the recent $w$ weeks. Given $D=dw$ days are used in total, the \emph{joint matrix} is represented as:
\begin{equation}
\begin{bmatrix}
  \mathbf{Y}^{tr}_D&\cdots  & \mathbf{Y}^{tr}_1  &\mathbf{Y}^{te} \\ 
\phi(\mathbf{X}^{tr}_D) &  \cdots& \phi(\mathbf{X}^{tr}_1)& \phi(\mathbf{X}^{te})
\end{bmatrix}\in\mathbb{R}^{(n+h)\times(DT+T_{te})}
\label{eq:19}
\end{equation}
For simplicity, the \emph{joint matrix} in \eqref{eq:19} is also written as 
\begin{equation}
\begin{bmatrix}
 \mathbf{Y}^{tr}&\mathbf{Y}^{te} \\ 
\phi(\mathbf{X}^{tr})&\phi(\mathbf{X}^{te}) 
\end{bmatrix}.
\end{equation}


\subsubsection{Ensembled Kernelized Matrix Completion (EKMC)}
 The basic idea behind EKMC is to iteratively solve a KMC problem for prediction, and combine the prediction results of each problem using a weighted majority strategy. Specifically, we assign each column of the training data $\mathbf{X}^{tr}(:,i) = \RX_1(:,i)$ a weight vector $\boldsymbol{\theta}(i)$. The weight vector is updated in each iteration according to the prediction accuracy of $\mathbf{Y}^{tr}(:,i)$, which makes the EKMC algorithm focus on hard-to-predict samples. Thus, their combination is expected to generate improved prediction results. The algorithm is summarized in Alg. \ref{A:1} below.
\begin{algorithm}[htbp] 
 \caption{EKMC Algorithm} 
 \label{A:1}
 \begin{algorithmic}[1]
\Require Joint matrix $\mathbf{Z}\in\mathbb{R}^{(n+h)\times(DT+T_{te})}$ in \eqref{eq:19};
\State \textbf{Initialize}: Weight vector $\boldsymbol{\theta}_{[0]}(i)=1,i=1,\cdots, DT$; 
\While {convergence criteria not met}

\State $\RX_1^{k+1}(:,i)= \boldsymbol{\theta}^k(i) \RX_1^k(:,i)$;

\State Estimate $[\tilde{\mathbf{Y}}^{te,k+1},\tilde{\mathbf{Y}}^{tr,k+1}]$ by KMC;

\State Compute the error: 

$\varepsilon^{k+1} = \frac{\sum_i \boldsymbol{\theta}^k(i) \mathbb{I}(\tilde{\mathbf{Y}}^{tr,k+1}(:,i) \neq \mathbf{Y}^{tr,k+1}(:,i))}{\sum_{i=1}^{DT} \boldsymbol{\theta}^k(i)}$;

\State Compute the update factor: 

$\beta^{k+1} = \log \frac{1-\varepsilon^{k+1}}{\varepsilon^{k+1}}$;

\State Weight update:

 $\boldsymbol{\theta}^{k+1}(i)=\boldsymbol{\theta}^{k+1}(i)e^{\beta^k \mathbb{I}(\tilde{\mathbf{Y}}^{tr,k+1}(:,i)\neq \mathbf{Y}^{tr,k+1}(:,i))}$;

\EndWhile

\State $\tilde{\mathbf{Y}}^{te}(:,i)=\sum_k \frac{\beta^k}{\sum_{j} \beta^j} \tilde{\mathbf{Y}}^{te,k}(:,i)$, for $i=1,\hdots, DT$;

\State $\tilde{\mathbf{Y}}^{te}(j,i)=\left\{\begin{matrix}
 1, &\tilde{\mathbf{Y}}^{te}(j,i)> 0.5 \\ 
 0,& \textrm{else}
\end{matrix}\right.$

\Ensure  $\tilde{\mathbf{Y}}^{te}$;
\end{algorithmic}
\end{algorithm}












The thresholding in step 10 of the algorithm above can be generalized to a column-specific threshold, which improves the performance of the algorithm.  We refer to \cite{li2020nonlinear} for example.  For the fixed threshold, it has been established that the training error after $K$ iterations is bounded by \cite{freund1997decision}
\begin{equation}
	2^K DT \prod_{k=1}^K \sqrt{\varepsilon^k(1-\varepsilon^k)}.
\end{equation}
%
This bound allows for estimating a number of iterations before running the algorithm.  Suppose $K$ is the number of iterations selected.  Then the time complexity of EKMC is $O(KD^2T^2r)$.


\section{Experimental Results}
\subsection{Data Description}
\begin{figure}[htbp]
\centerline{\includegraphics[scale=0.5]{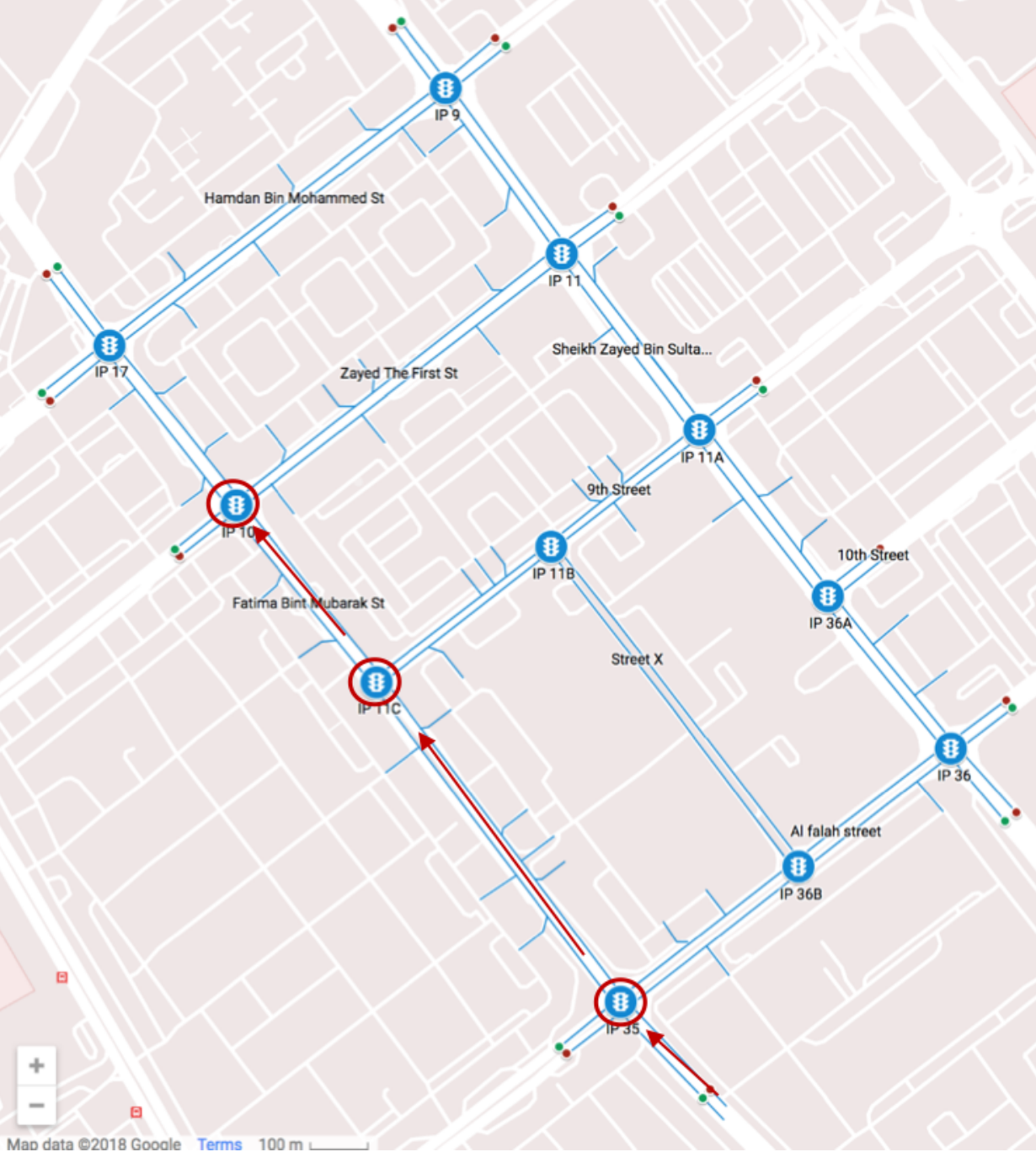}}
		\caption{Real world network in Abu Dhabi, UAE\label{F:4}}		
\end{figure}

High-resolution data are obtained from Al Zahiyah in downtown Abu Dhabi, UAE.  This traffic network consists of 11 signalized intersections and two parallel major arterials, which are presented in Fig. \ref{F:4}. We select data from three sensors located at three adjacent intersections marked red in the figure, where the circles represent the selected intersections and arrows represent the direction of travel.  
We test the proposed method using seven weeks of data (49 days, from the beginning of the 1st week in December, 2018 to the end of the 3rd week in January, 2019). It is worth noting that workdays and weekends are not distinguished in our experiments as the proposed method is essentially a dynamic learning approach that is adaptive to time-varying changes.

\subsection{Experimental Setup}
The parameters of the problem are as follows:
\begin{enumerate}
\item We test with different prediction horizons, namely $H \in \{1,10,60,120\}$ seconds.
\item The search grid for parameter $T$ is $\{10,60,120,300\}$. The parameter $D=5 \times 4$, which represents 5 days and 4 weeks.
\end{enumerate}
 
We compare the proposed method with the following baseline algorithms: Vector Autoregressive (VAR), Support Vector Regression (SVR) and Recurrent Neural Network (RNN) with long short-term memory (LSTM) structure.  All the tests are run on a 2.7 GHz intel Core i7 Processor with 16 GB of RAM.

\subsection{Performance Indices}
In this work, two indexes are utilized to evaluated the overall performance of traffic prediction. The first is mean absolution error (MAE), which is computed as
\begin{equation}
\textrm{MAE}_j=\frac{1}{t}\sum_{i=1}^t |\mathbf{Y}_{i,j}-\tilde{\mathbf{Y}}_{i,j}|
\label{eq:20}
\end{equation}
where $\mathbf{Y}_{i,j}$ denotes the actual traffic data for variable $j$ and observation $i$, $\tilde{\mathbf{Y}}_{i,j}$ is its corresponding prediction and $t$ is number of observations.  Specifically, $\mathbf{Y}_{i,j} \in \{0,1\}$ is sensor $j$'s occupancy at time $i$ seconds, where `1' represent an occupied state and `0' represents an unoccupied sensor. 
The second index is the \emph{Skorokhod} $M_1$ metric \cite{whitt2002stochastic}, which is defined as 
\begin{equation}
d_{M_1}(\mathbf{Y}_j,\tilde{\mathbf{Y}}_j)\equiv \underset{\substack{(u_j,r_j)\in\Pi(\mathbf{Y}_j)\\ (\tilde{u}_j,\tilde{r}_j)\in\Pi(\tilde{\mathbf{Y}}_j)}}{\textrm{inf}} \{\sup|u_j-\tilde{u}_j|\vee \sup|r_j-\tilde{r}_j|\}
\label{eq:21}
\end{equation}
where $a\vee b\equiv \textrm{max}\{a,b\}$, $\Pi(\mathbf{Y}_j)$ and $\Pi(\tilde{\mathbf{Y}}_j)$ are respectively the sets of parametric representation of $\mathbf{Y}_j = \{\mathbf{Y}_{i,j}\}_{i=1}^t$ and $\tilde{\mathbf{Y}}_j = \{\tilde{\mathbf{Y}}_{i,j}\}_{i=1}^t$. $\sup|u_j-\tilde{u}_j|$ measures the sensor state difference and $\sup |r_j-\tilde{r}_j|$  measures the time shift introduced by the parametric representations of the two signals.  In essence, the parametric representations \emph{shift} the two signals $\mathbf{Y}_j$ and $\tilde{\mathbf{Y}}_j$ in a way that they align best (via the infimum operation) but that shift is penalized by the metric: the first term $\sup|u_j-\tilde{u}_j|$ cannot exceed 1 and, hence, a shift that exceeds 1 second will render the second term $\sup |r_j-\tilde{r}_j|$ will dominate.

\subsection{Results and Discussion}
The result of 1-MAE for the proposed method w.r.t. $H$ and $DT$ are summarized in Table \ref{table_2}, from which we can see that the prediction accuracy of the proposed method can exceed $92\%$ but does not fall below $75\%$. Note that the prediction accuracy decreases as $H$ increases, and with the increase of historical data, the prediction accuracy first increases and then drops.  Historical data contain useful information that can benefit prediction but for high resolution contexts, this seems to diminish with more historical data as irrelevant information impacts performance.  As seen in Table \ref{table_2}, the best overall accuracy is observed at $H=1$ and $DT=1200$ for all three sensors. 
\begin{table}[thbp]
	\caption{Prediction Performance of the proposed method regarding different parameters}
	\centering
\subtable[Sensor 1]{\begin{tabular}{ccccc}
\hline\hline \\[-3mm]
 \diagbox{DT}{PH} & {1} & {10} & {60} & {120}
                                                                                \\[1.6ex] \hline
200            &0.8848            &  0.8611   &0.8415& 0.8018                                                                             \\
\textbf{1200}             & \textbf{0.9067}        & \textbf{0.8867}   &\textbf{0.8495} &\textbf{0.8275}                                                                              \\
2400 &0.8719 &0.8487 & 0.8178&0.8025 \\
6000 &0.8455&0.8215&0.8055&0.7818\\
\hline
\hline
\end{tabular}
\label{table_2a}
}
\qquad
\subtable[Sensor 2]{\begin{tabular}{ccccc}
\hline\hline \\[-3mm]
 \diagbox{DT}{H} & {1} & {10} & {60} & {120}
                                                                                \\[1.6ex] \hline
200            &0.9045            &0.8820     &0.8380 &    0.7878                                                                          \\
\textbf{1200}             & \textbf{0.9203}       &  \textbf{0.9050}  & 0.\textbf{8500} &     \textbf{ 0.8215}                                                                        \\
2400 &0.8638& 0.8355&0.8028 &0.7548\\
6000 &0.8358& 0.8078&0.7815&0.7505\\
\hline
\hline
\end{tabular}
\label{table_2b}
}
\qquad
\subtable[Sensor 3]{\begin{tabular}{ccccc}
\hline\hline \\[-3mm]
 \diagbox{DT}{PH} & {1} & {10} & {60} & {120}
                                                                                \\[1.6ex] \hline
200            & 0.8846          &  0.8648    &\textbf{0.8550}&         0.8150                                                                     \\
\textbf{1200}             &     \textbf{0.8950}    &  \textbf{0.8689}  &0.8484&     0.8300                                                                        \\
2400 &0.8800&0.8620&0.8327&\textbf{0.8309}\\
6000 &0.8550&0.8348&0.8288&0.8019\\
\hline
\hline
\end{tabular}
}
\label{table_2}
\end{table}
Table \ref{table_3} shows the results of 1-$d_{M_1}$ for each sensor with parameters $DT=1200$ and $H=1,10,60,120$. It can be observed that the resulting accuracies are no less than 0.82 with highest one exceeding 0.90. 
\begin{table}[thbp]
	\caption{Prediction performance of the proposed method evaluated by 1-$d_{M_1}$}
	\centering
{\begin{tabular}{ccccc}
\hline\hline \\[-3mm]
 \diagbox{PH}{Sensor} & {1} & {2} & {3} & {$\textrm{Mean}$}
                                                                                \\[1.6ex] \hline
1            &0.8755          &  0.9089   &0.8533   &   0.8792                                                                  \\
10          & 0.8640     &   0.8848 &   0.8615      &   0.8701                                                         \\
60          &  0.8309    &   0.8513 &   0.8489     &     0.8437                                                        \\
120          &  0.8255    &   0.8235 &   0.8350     &     0.8280                                                           \\
\hline
\hline
\end{tabular}
\label{table_3}
}
\end{table}


We further compare the prediction performance of the proposed method (EKMC) and other baseline algorithms. As shown in Table \ref{table_4}, the proposed EKMC outperforms VAR and SVR significantly, and achieves comparable performance as RNN with basic LSTM; the overall prediction accuracy of RNN is a little higher than EKMC (1-MAE) but EKMC shows better pattern recognition capabilities (1-$d_{M_1}$).
\begin{table}[thbp]
	\caption{Prediction performance of the proposed method and other baseline algorithms}
	\centering
{\begin{tabular}{ccc}
\hline\hline \\[-3mm]
 \diagbox{Method}{Index} & { 1-$d_{M_1}$ }  & {1-MAE}
                                                                                \\[1.6ex] \hline
MC                   &0.8105&0.8674                            \\                                
EKMC            &0.8792             &   0.9073                                                                   \\
VAR            & \textrm{NAN}             & 0.8014                                                                \\
SVR &0.7805&0.8669\\
RNN  &0.8753&0.9151\\
\hline
\hline
\end{tabular}
\label{table_4}
}
\end{table}
\begin{figure}[htbp]
\centering
{\includegraphics[scale=0.45]{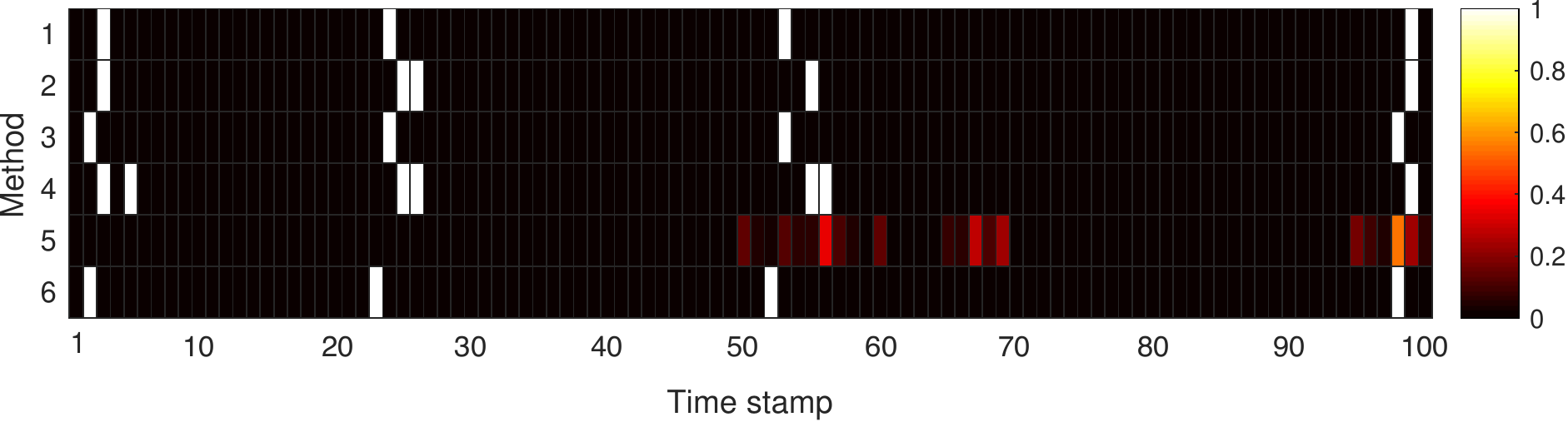}\label{F:7-1}}
		\caption{Prediction results of (a) variable 1 and (b) variable 3 based on different methods (1 = EKMC, 2 = MC, 3 = RNN, 4 = SVR, 5 = VAR, 6 = Ground Truth)\label{F:7}}		
\end{figure}

Fig. \ref{F:7} presents an example of the prediction results of different methods w.r.t. sensor 1. As shown in Fig. \ref{F:7}, sensor 1 has four jumps in the ground truth series (which occur at 2s, 23s, 52s and 98s). The closest prediction results are generated by RNN, where jumps are predicted as $\{2,\ 21,\ 97,\ 98\}$ which almost captures the truth except the jump at time 52 seconds. This is followed by the proposed EKMC algorithm which predicts the jumps at $\{3,\ 24,\ 54,\ 99\}$. Note that the overall prediction accuracy (evaluated by 1-MAE) of RNN is higher than EKMC while the jump patterns (evaluated by 1-$d_{M_1}$) were better captured by EKMC. SVR captures a rough pattern of jumps with incorrect number of total jumps. As for the VAR, it only tends to capture the mean trend in the high-resolution data. 

\begin{figure}[htbp]
\centering
{\includegraphics[scale=0.55]{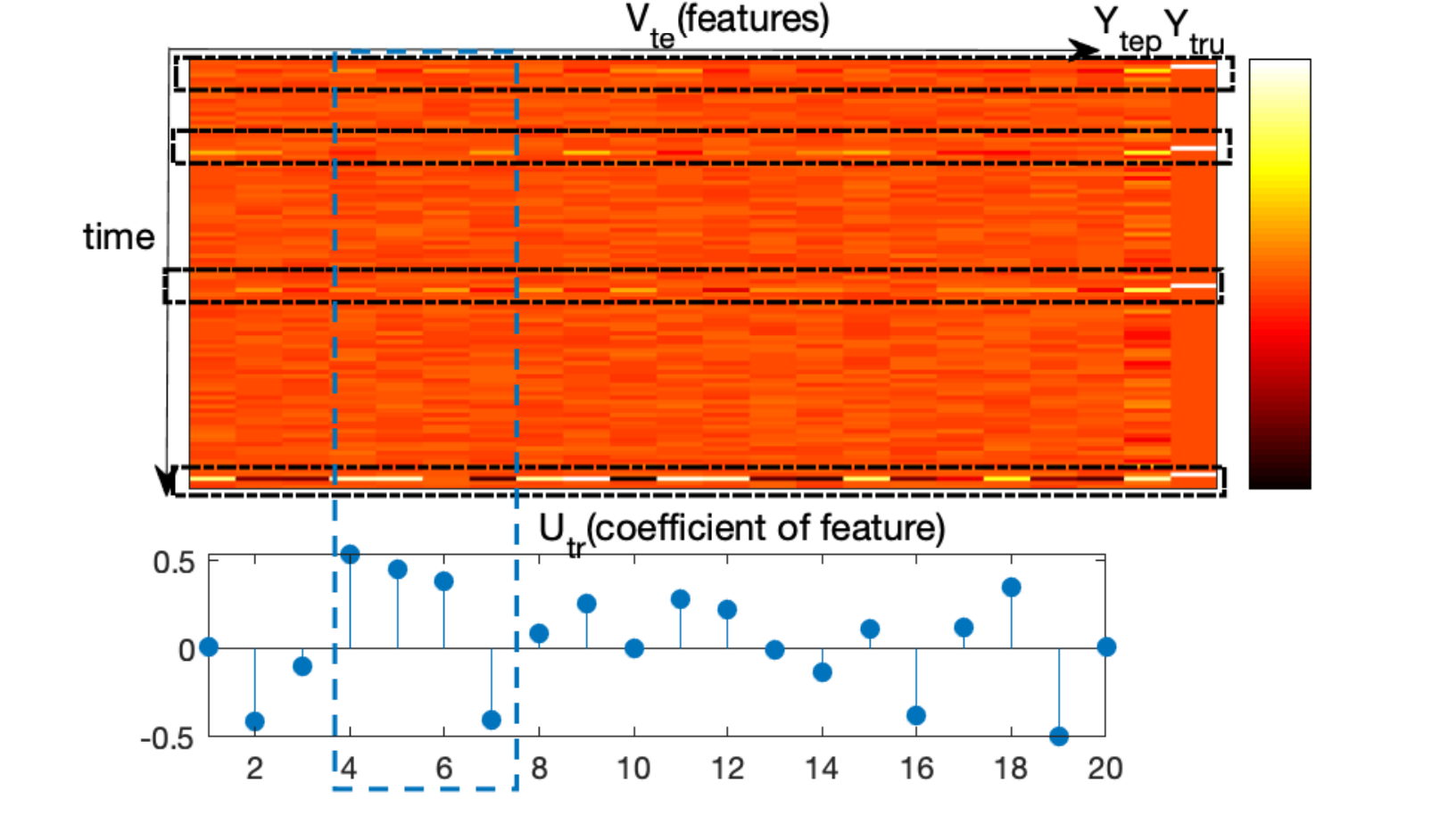}
		\caption{An Illustration of interpretability of features from EKMC\label{F:8}}
		}		
\end{figure}

Another advantage of the proposed is \emph{interpretability} of the results.  This is illustrated by the following example in Fig. \ref{F:8} (For simplicity, we adopt the same case shown in Fig. \ref{F:7}), where the top sub-figure is the heatmap of the matrix consisting of features  $\mathbf{V}_{te}$ (each column corresponds to a feature vector, and we have 20 features), predictions  $\mathbf{Y}_{tep}$ and the ground truth  $\mathbf{Y}_{tru}$ w.r.t sensor 1. The bottom sub-figure shows the coefficients $\mathbf{U}_{tr}$ of features  (it is noted that $\mathbf{Y}_{tep}=\mathbf{V}_{te}\mathbf{U}_{tr}^\top$). We can see from the heatmap figure that, despite a small time delay, the features associated with the time points with jumps (which are marked with black rectangles) are different from those corresponding to non-jumps. Moreover, we observe that different features weigh differently during the calculation of prediction, i.e., features 4, 5, 6 are quite similar to the ground truth (they have similar colors) and might well represent the truth, thus their coefficients are large. On the other hand, the colors associated with feature 7 differ from the ground truth, which implies that feature 7 may not be a good predictor. Thus, feature 7 has a very small (negative) coefficient.

\section{Conclusion}
We propose a novel traffic prediction method based on high-resolution data, termed \emph{ensembled kernelized matrix completion} (EKMC). The kenelized settings allows us to learn nonlinear dependencies in high-resolution traffic data and an ensemble learning strategy ensures high prediction accuracy over all the data (not only the majority of the data). The KMC algorithm has good convergence properties (i.e., sub-linearly convergence) and EKMC has a bounded training error. We conduct extensive experiments based on real world high-resolution data from downtown Abu Dhabi, where the results showcase that the proposed EKMC works well for high-resolution data prediction, it outperforms VAR and SVR and even achieves comparable prediction performance with the powerful RNN with LSTM network.  However, in contrast to the latter, our proposed EKMC is an ``\emph{open-box}'' tool in that it offers simple interpretability features that cannot be obtained with neural networks.

\bibliographystyle{IEEEtran}
\bibliography{IEEEexample}
\vspace{12pt}

%



\end{document}